\documentclass{PoS}

\pdfoutput=1

\usepackage[utf8]{inputenc}
\usepackage{amsmath}
\usepackage{url}
\usepackage{xspace}

\newcommand{\Pythia}{\textsc{Pythia}}

\newcommand{\UNIT}[1]{\ensuremath{\,{\rm #1}}\xspace}

\newcommand{\keV}{\UNIT{keV}}
\newcommand{\MeV}{\UNIT{MeV}}
\newcommand{\GeV}{\UNIT{GeV}}

\newcommand{\fm}{\UNIT{fm}}

\let\oldsqrt\sqrt
\def\sqrt{\mathpalette\DHLhksqrt}
\def\DHLhksqrt#1#2{%
\setbox0=\hbox{$#1\oldsqrt{#2\,}$}\dimen0=\ht0
\advance\dimen0-0.2\ht0
\setbox2=\hbox{\vrule height\ht0 depth -\dimen0}%
{\box0\lower0.4pt\box2}}

\title{In-Medium Properties of Vector Mesons in a Transport Approach}
\ShortTitle{In-Medium Properties of Vector Mesons in a Transport Approach}

\author{\speaker{Janus Weil}\\
        Institut f\"ur Theoretische Physik, Universit\"at Giessen, Germany\\
        E-mail: \email{janus.weil@theo.physik.uni-giessen.de}}

\author{Kai Gallmeister\\
        Institut f\"ur Theoretische Physik, Universit\"at Giessen, Germany\\
        E-mail: \email{kai.gallmeister@theo.physik.uni-giessen.de}}

\author{Ulrich Mosel\\
        Institut f\"ur Theoretische Physik, Universit\"at Giessen, Germany\\
        E-mail: \email{ulrich.mosel@theo.physik.uni-giessen.de}}

\abstract{We present dilepton spectra from p+p and p+Nb collisions at a kinetic
beam energy of 3.5 \GeV, which were simulated with the GiBUU transport model
assuming different in-medium scenarios. We compare these spectra to preliminary
HADES data and show that GiBUU can describe the data reasonably well. Our
simulations indicate that the intermediate dilepton-mass region is sensitive to
the N-$\Delta$ electromagnetic transition form factor, which up to now is
unmeasured in the time-like region.}

\FullConference{ XLIX International Winter Meeting on Nuclear Physics\\
		 24-28 January 2011\\
		 BORMIO, Italy}

\begin{document}

\parindent0cm

%%%%%%%%%%%%%%%%%%%%%%%%%%%%%%%%%%%%%%%%%%%%%%%%%%%%%%%%%%%%%%%%%%%%%%%%%%%%%%%

\section{Introduction}

While the vacuum properties of most hadrons are known to reasonable
accuracy nowadays, it is a heavily debated question how these
properties change inside nuclear matter. In particular, for the light
vector mesons, various theoretical predictions
regarding their in-medium properties have been suggested. 

Among these expected in-medium effects, a so-called ``collisional
broadening'' of the meson spectral function, due to inelastic
collisions with the hadronic medium, is presumed.
A second class of predictions claims that the vector meson masses
will be shifted in the medium. These changes of the peak mass are
connected to the partial restoration of chiral symmetry in the medium
and have been studied via QCD sum rules. This effect has been claimed
to be seen in experiments, but is still being discussed
controversially. For a recent review on in-medium effects, see
\cite{Leupold:2009kz}. 

For studying in-medium effects, the more prominent hadronic decay
modes of the vector mesons are unfavorable, since they are affected by
strong final-state interactions with the hadronic medium -- in
contrast to the rare dilepton decay modes, which only feel the
electromagnetic force. Therefore the latter are ideally suited to
carry the in-medium information outside to the detector, undisturbed
by the hadronic medium. 

Experimentally, dilepton spectra from elementary nuclear reactions are
being studied for example with the CLAS detector at JLAB, where photons with
energies of a few \GeV interact with nuclei \cite{Wood:2008ee}, or by the
E325 experiment at KEK, where 12 \GeV protons are used as projectile
\cite{Naruki:2005kd}. Also, the HADES detector at GSI has an ambitious
program for measuring dilepton spectra from p+p, p+A and A+A reactions
\cite{:2008yh}. On the side of the hadronic decays, most notably
$\omega\rightarrow\pi^0\gamma$ is being investigated by the CB/TAPS
group in photon-induced reactions at the ELSA accelerator
\cite{Nanova:2010tq}. 

In this paper, we apply the Gießen Boltzmann-Uehling-Uhlenbeck
transport model (GiBUU) \cite{gibuu} to the p+p and p+Nb reactions
studied by the HADES collaboration. 
We use GiBUU to generate dilepton events and pass them through the
HADES acceptance filter, in order to compare our calculations directly
to the experimental data measured by HADES.   

%%%%%%%%%%%%%%%%%%%%%%%%%%%%%%%%%%%%%%%%%%%%%%%%%%%%%%%%%%%%%%%%%%%%%%%%%%%%%%%

\section{The GiBUU Transport Model}

\label{sec:gibuu}

Our tool for the numerical simulation of dilepton spectra is the
GiBUU hadronic transport model, which provides a unified
framework for various types of elementary reactions on nuclei as well
as heavy-ion collisions \cite{gibuu}. This model takes care
of the correct transport-theoretical description of the hadronic
degrees of freedom in nuclear reactions, including the propagation,
elastic and inelastic collisions and decays of particles.
The GiBUU model is based on the Boltzmann-Uehling-Uhlenbeck equation,

\begin{equation}
 ( \partial_t + \vec\nabla_p H\cdot\vec\nabla_r - \vec\nabla_r
 H\cdot\vec\nabla_p ) f_i(\vec r,p,t) = I_{\rm coll} [f_i,f_j,...] \; ,
\end{equation}

which describes the space-time evolution of the one-particle
phase-space density $f_i$ of a given particle species 
$i$ under the influence of a mean-field potential.
The phase-space densities $f_i$ at time $t$ depend on the spatial coordinates
$\vec r$ and the four momentum $p$. The right-hand side
of the equation is given by the collision term $I_{\rm coll}$, which
describes collisions and decays of particles. The left-hand side,
the so-called ``Vlasov'' part,
describes the propagation of particles in a mean field, where $H$
is the relativistic one-particle Hamiltonian. In order to solve the
BUU equation numerically, we rely on the test-particle ansatz. Here
the phase-space densities are approximated by a large number of
test particles, each represented by a $\delta$-distribution in
coordinate and momentum space. 

The propagation of particles with density-dependent spectral functions
(usually referred to as ``off-shell propagation'') poses a particular challenge.
Our approach to this problem is based on the off-shell equations of motion of
test particles, as given in \cite{Cassing:1999wx}. Such an off-shell treatment
is necessary for including in-medium modifications of the spectral functions
(e.g. collisional broadening of the vector mesons). The collisional width inside
a nuclear medium of density $\rho$ can be related to the collision cross section
$\sigma_{NX}$ with the nucleons in low-density approximation as

\begin{equation}
 \Gamma_{\rm coll} = \rho \left< v_{\rm rel} \sigma_{NX} \right> \; ,
\end{equation}

where $v_{rel}$ is the relative velocity and the brackets indicate an
integration over the Fermi momentum of the nucleons. In general this collisional
width will depend on the momentum of the involved particle $X$. However, a
consistent treatment of such a momentum-dependent collisional width is not
possible with the off-shell equations of motion mentioned above, as it will lead
to superluminous test particles. Therefore we have to neglect the momentum
dependence and keep only the linear density dependence

\begin{equation}
 \Gamma_{\rm coll} = \Gamma_0 \cdot \frac{\rho}{\rho_0} \; ,
\end{equation}

where $\rho_0=0.168\fm^{-3}$ is the normal nuclear matter density. The value of
$\Gamma_0$ should on average match the momentum-dependent width as obtained from
the collision term. For the $\rho$ and $\omega$ mesons we typically use
$\Gamma_0 = 150 \MeV$.

%%%%%%%%%%%%%%%%%%%%%%%%%%%%%%%%%%%%%%%%%%%%%%%%%%%%%%%%%%%%%%%%%%%%%%%%%%%%%%%

\section{The Collision Term / Elementary Cross Sections}

The collision term contains all sorts of 
scattering and decay processes: elastic and inelastic two-body collisions,
decays of unstable resonances and even three-body collisions (which are only
relevant at high densities).

The two-body part of the collision term is separated into two different regimes
in terms of the available energy $\sqrt{s}$:
a resonance model description at low energies and the \Pythia{}
string model at high energies. 

For baryon-baryon collisions, the transition between the two is usually 
performed at $\sqrt{s}=2.6\GeV$. There is a small window
around this border ($\pm0.2\GeV$), where both models are faded
linearly into each other in order to ensure a smooth transition. 
For meson-baryon collisions, the transition region lies at
$\sqrt{s}=2.2\pm0.2\GeV$.  

The low-energy part is given by a resonance model \cite{Teis:1996kx},
where basically all collision cross sections are assumed to be
dominated by the excitation of nucleon resonances. The GiBUU model
currently contains 61 baryon species (around 30 nucleon resonances
plus strange and charmed baryons). The resonance model gives a good
description of exclusive channels like one- and two-pion production as
well as single $\eta$ and $\rho$ production. 

At higher energies the resonance model breaks down. There we
rely on the Monte Carlo event generator \Pythia{} (v6.4.24)
\cite{pythia,Sjostrand:2006za}, which is based on the Lund string model.
Although \Pythia{} clearly has its strengths in the high-energy regime,
we use it down to energies of a few \GeV. This works surprisingly
well, as has recently been demonstrated for example by GiBUU's
successful description of pion data measured by the HARP collaboration
\cite{Gallmeister:2009ht}. In the following, we will show that our approach also
provides a rather good description of HADES dilepton spectra. 

Since GiBUU is intended to be a multi-purpose generator, we follow the
philosophy that \Pythia{} is used with most of its default settings.
It is obvious, however, that tuning a few selected parameters can lead to a much
better agreement with specific experimental data; cf.~appendix A for
the parameter set used for the HADES calculations. 

Furthermore, we had to extend \Pythia{} in some more aspects. For
instance \Pythia{}'s treatment of spectral functions is rather
simplified and not sufficient for a proper description of dilepton
spectra, which are very sensitive to the vector meson
spectral functions. Therefore we had to replace \Pythia{}'s internal spectral
functions by proper ones, because the former lack mass-dependent decay widths
and are restricted to a narrow region around the pole mass. This also gives us
the opportunity to introduce density-dependent spectral functions. 

%%%%%%%%%%%%%%%%%%%%%%%%%%%%%%%%%%%%%%%%%%%%%%%%%%%%%%%%%%%%%%%%%%%%%%%%%%%%%%%

\section{Dilepton Decays and Form Factors}

In the GiBUU model the following dilepton decay modes are taken into account:
\begin{itemize}
\item direct decays, as $V \rightarrow e^+e^-$ with
  $V=\rho^0,\omega,\phi$\quad or\quad $\eta \rightarrow e^+e^-$\quad,
\item Dalitz decays, as $P \rightarrow e^+e^-\gamma$ with
  $P=\pi^0,\eta,\eta'$\quad or\quad $\omega \rightarrow \pi^0e^+e^-$\quad or
  \quad $\Delta \rightarrow Ne^+e^-$\quad.
\end{itemize}
Most of them are
treated similarly as in \cite{effe_phd}. The leptonic decay widths of
the vector mesons are taken under the assumption of strict
vector-meson dominance (VMD),
\begin{equation}
  \Gamma_{V\rightarrow e^+e^-}(\mu)=C_V\frac{m_V^4}{\mu^3},
\end{equation}
with the constants $C_V$ listed in table \ref{tab:vm_dil} (taken from
\cite{Nakamura:2010zzi}).

\begin{table}[h]
  \begin{center}
    \begin{tabular}{|c|c|c|c|c|c|}
      \hline
      $V$ & $m_V (\MeV)$ & $\Gamma_{ee} (\keV)$ & $C_V=\Gamma_{ee}/m_V$ \\
      \hline
      $\rho$   & 775.49   & 7.04 & $9.078\cdot10^{-6}$ \\
      $\omega$ & 782.65   & 0.60 & $7.666\cdot10^{-7}$ \\
      $\phi$   & 1019.455 & 1.27 & $1.246\cdot10^{-6}$\\
      \hline
    \end{tabular}
  \end{center}
  \caption{Dilepton-decay constants for $V\rightarrow e^+e^-$.}
  \label{tab:vm_dil}
\end{table}

While the direct decay of the $\eta$ meson into a $\mu^+\mu^-$ pair
has been observed, for the corresponding $e^+e^-$ decay only an upper
limit of $\mbox{BR}(\eta\rightarrow e^+e^-)<2.7\cdot10^{-5}$ is known
\cite{Berlowski:2008zz}. However, the theoretical expectation from
helicity suppression is still four orders of magnitude lower
\cite{Browder:1997eu}.

The Dalitz decays of the pseudoscalar mesons, $P=\pi^0,\eta,\eta'$,
are treated via the parametrization \cite{Landsberg:1986fd},
\begin{align}
  \frac{d\Gamma_{P\rightarrow\gamma e^+e^-}}{d\mu} = &
  \frac{4\alpha}{3\pi}\frac{\Gamma_{P\rightarrow\gamma\gamma}}{\mu}
  \left(1-\frac{\mu^2}{m_P^2}\right)^3 |F_P(\mu)|^2,
\end{align}
with $\Gamma_{\pi^0\rightarrow\gamma\gamma}=7.8\cdot10^{-6}\MeV$,
$\Gamma_{\eta\rightarrow\gamma\gamma}=4.6\cdot10^{-4}\MeV$ and the
form factors
\begin{alignat}{4}
  F_{\pi^0}(\mu) & = 1 + b_{\pi^0}\mu^2, \quad & b_{\pi^0} & = 5.5\GeV^{-2}, \\
  F_{\eta}(\mu) & = \left(1-\frac{\mu^2}{\Lambda_\eta^2}\right)^{-1},
  \quad & \Lambda_\eta & = 0.676\GeV\quad.
\end{alignat}

The above value of $\Lambda_\eta$ has been recently determined
from the HADES data at 2.2 \GeV beam energy \cite{spruck_phd}. It
should be noted that the form factors of the $\pi^0$ and $\eta$
Dalitz decays are sufficiently constrained by data, while the
experimental constraints of the $\eta'$ form factor are much weaker
\cite{Landsberg:1986fd}. A VMD form factor for the $\eta'$ Dalitz
decay can be found for example in \cite{Terschluesen:2010ik}. However, the
$\eta'$ contribution to the HADES dilepton spectra turns out to be
practically insignificant. The parametrization of the $\omega$ Dalitz
decay

\begin{align}
  \frac{d\Gamma_{\omega\rightarrow\pi^0e^+e^-}}{d\mu} = & \frac{2\alpha}{3\pi}\frac{\Gamma_{\omega\rightarrow\pi^0\gamma}}{\mu}
  \left[ \left(1+\frac{\mu^2}{\mu_\omega^2-m_\pi^2}\right)^2 -\frac{4\mu_\omega^2\mu^2}{(\mu_\omega^2-m_\pi^2)^2} \right]^{3/2}
  |F_\omega(\mu)|^2, \\
  |F_\omega(\mu)|^2 = & \frac{\Lambda_\omega^4}{(\Lambda_\omega^2-\mu^2)^2+\Lambda_\omega^2\Gamma_\omega^2},
\end{align}

is adopted from \cite{Bratkovskaya:1996qe,effe_phd} with
$\Gamma_{\omega\rightarrow\pi^0\gamma}=0.703\MeV$,
$\Lambda_\omega=0.65\GeV$ and $\Gamma_\omega=75\MeV$.  We note here
that the form factor of the $\omega$ Dalitz decay is also
well-constrained by data \cite{:2009wb}.

For the $\Delta$-Dalitz decay, we use the parametrization from \cite{Krivoruchenko:2001hs},

\begin{align}
  \frac{d\Gamma_{\Delta\rightarrow Ne^+e^-}}{d\mu} = & \frac{2\alpha}{3\pi\mu}\Gamma_{\Delta\rightarrow N\gamma^*}, \\
  \Gamma_{\Delta\rightarrow N\gamma^*} = &
  \frac{\alpha}{16}\frac{(m_\Delta+m_N)^2}{m_\Delta^3m_N^2}
  \left[(m_\Delta+m_N)^2-\mu^2\right]^{1/2}
  \left[(m_\Delta-m_N)^2-\mu^2\right]^{3/2} |F_\Delta(\mu)|^2,
\end{align}

where we neglect the electron mass. The electromagnetic N-$\Delta$ transition
form factor $F_\Delta(\mu)$ is an issue of
ongoing debate. Unlike the other semileptonic Dalitz decays, it is
poorly constrained by data. At least at the real-photon point
($\mu=0$) it is fixed by the decay $\Delta\rightarrow N\gamma$ to
$|F_\Delta(0)|=3.029$, and also in the space-like region this form
factor is well-constrained by electron scattering data on the
nucleon. However, this form factor is basically unknown in the time-like regime,
which is being probed by the $\Delta$ Dalitz decay.

From the theoretical side, many parametrizations are available for the
space-like part, but most of them are not applicable in the time-like
region. One of the few models which take care of the continuation to
the time-like region is the two-component quark model given in
\cite{Wan:2005ds}.

%%%%%%%%%%%%%%%%%%%%%%%%%%%%%%%%%%%%%%%%%%%%%%%%%%%%%%%%%%%%%%%%%%%%%%%%%%%%%%%

\section{Dilepton Spectra from Elementary p+p Collisions}

It is very important to make sure that one understands the elementary
reactions before moving on to heavier systems, which involve effects
of the nuclear medium. Fortunately, HADES has also measured dilepton
spectra from elementary p+p reactions. These provide a base line for
exploring the heavier nuclear systems.

\begin{figure}[ht]
  \begin{center}
  \includegraphics[height=5.1cm]{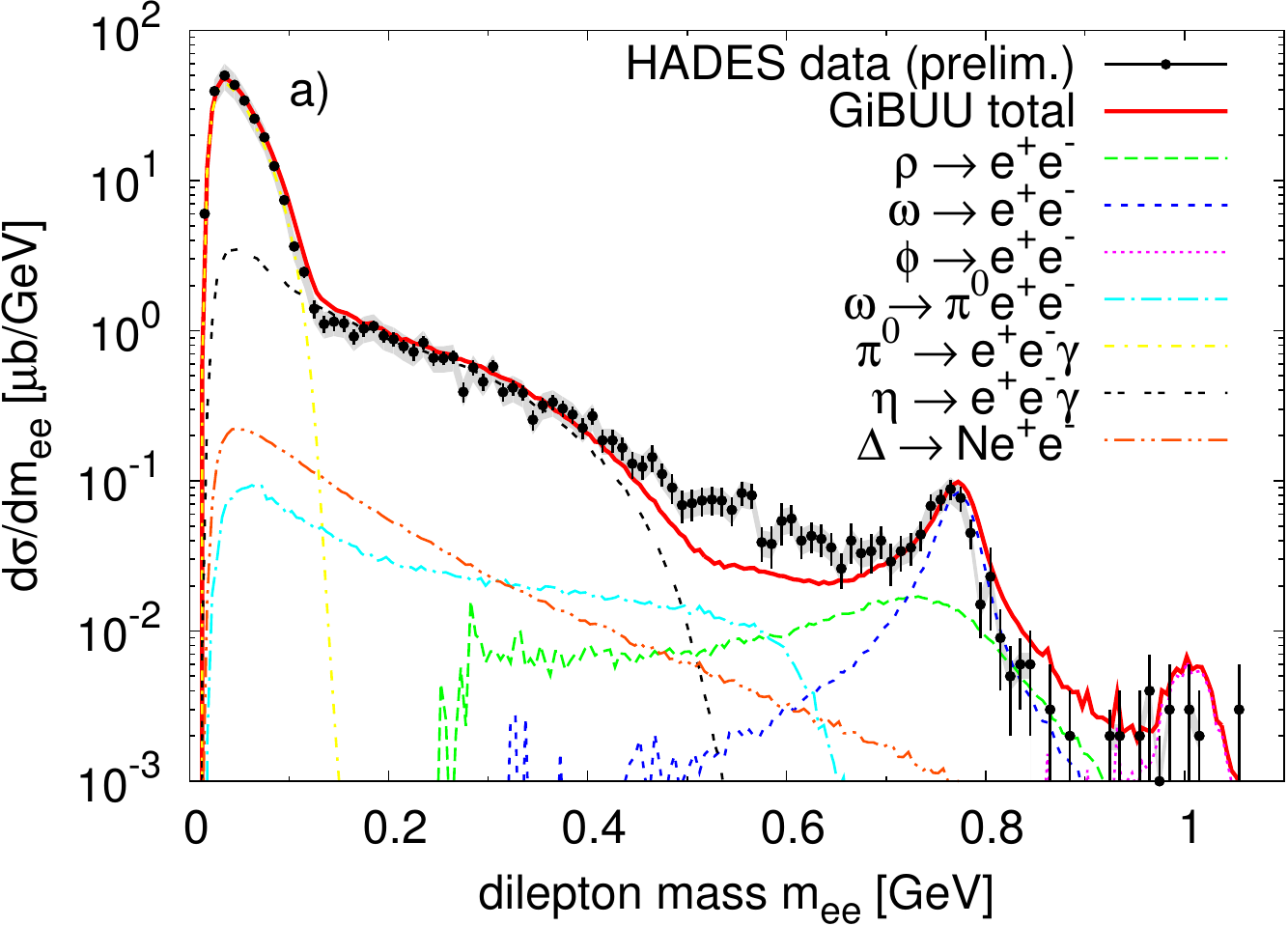}
  \includegraphics[height=5.1cm]{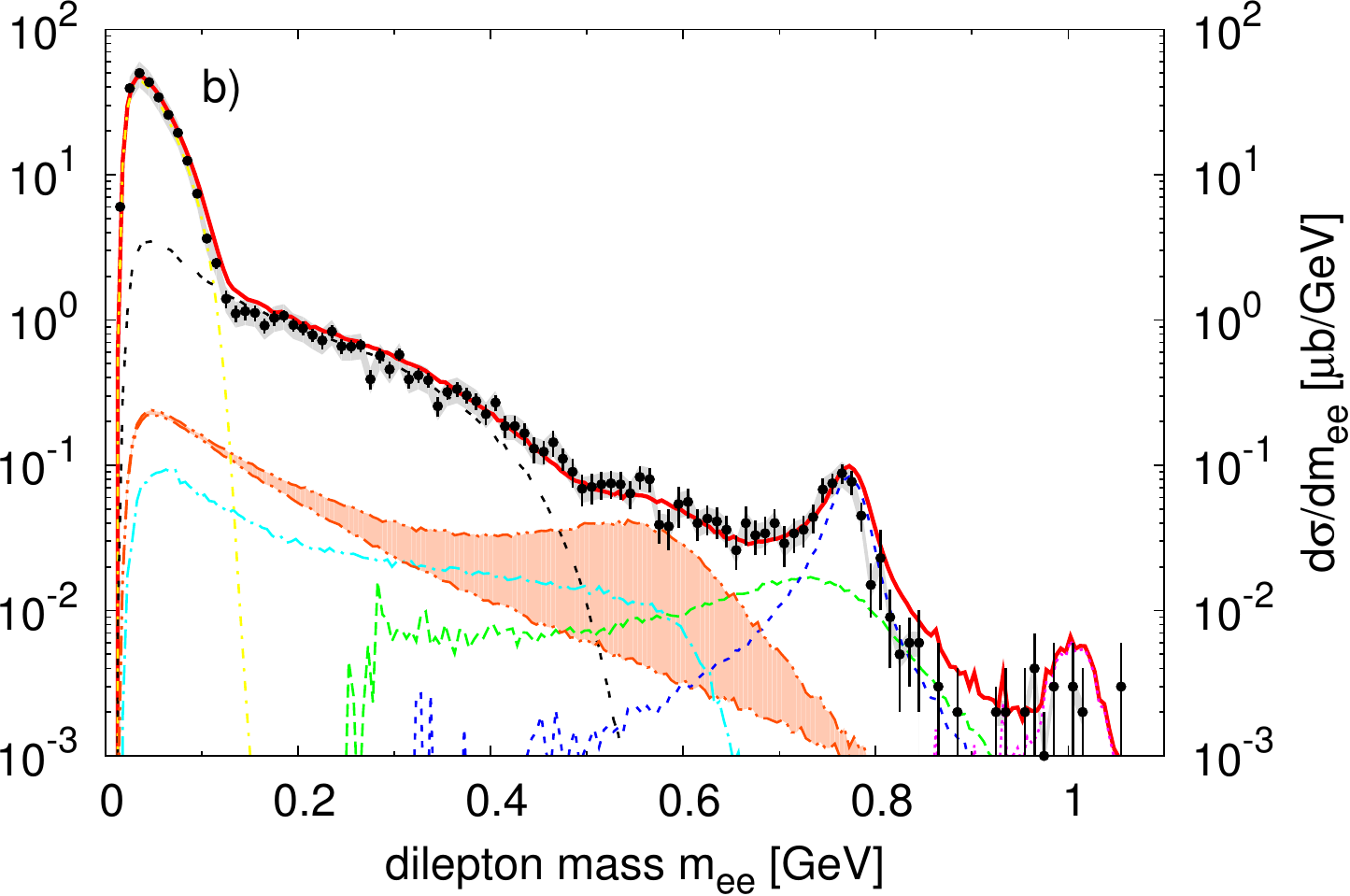}
  \end{center}
  \caption{(Color online) Dilepton spectra for p+p@3.5\GeV. a) Mass spectrum
    without $\Delta$ form factor, b) mass spectrum with $\Delta$ form factor.}
  \label{fig:pp_mass}
\end{figure}

Fig.~\ref{fig:pp_mass} shows a comparison plot of a GiBUU simulation to HADES
data \cite{Rustamov:2010zz} for a proton beam of 3.5 \GeV
kinetic energy impinging on a fixed proton target. This setup
corresponds to a center-of-mass energy of $\sqrt{s}=3.18\GeV$. The
theoretical results have been corrected for the HADES acceptance and
reasonably reproduce the shape of the data over most of the mass
spectrum. In the intermediate mass region around 500-600 \MeV it
seems like the inclusion of a proper transition form factor for the
$\Delta$ Dalitz decay \cite{Wan:2005ds} (shaded area) is crucial for
describing the data. Without such a form factor, the calculation
strongly underestimates the experimental data in this region (by at
least a factor of two). Another channel which could possibly contribute in
the intermediate mass region is $\eta\rightarrow e^+e^-$. The current upper
limit for the branching ratio of this decay would overshoot the HADES data
by at least a factor of four. However, there is no $\eta$ peak visible in the
data, and also the theoretical expectations from helicity suppression are still
orders of magnitude below the current experimental limit. Therefore it is rather
unlikely that the direct $\eta$ decay would give significant contributions to
the HADES dilepton spectrum at 3.5 \GeV.

It should be noted that the default \Pythia{}
parameters already give decent $\pi^0$ and $\eta$ production cross
sections, while the vector meson production is overestimated quite a
bit. This is cured by using the tuned parameters from appendix A.

\begin{figure}[ht]
  \begin{center}
    \includegraphics[width=0.9\textwidth]{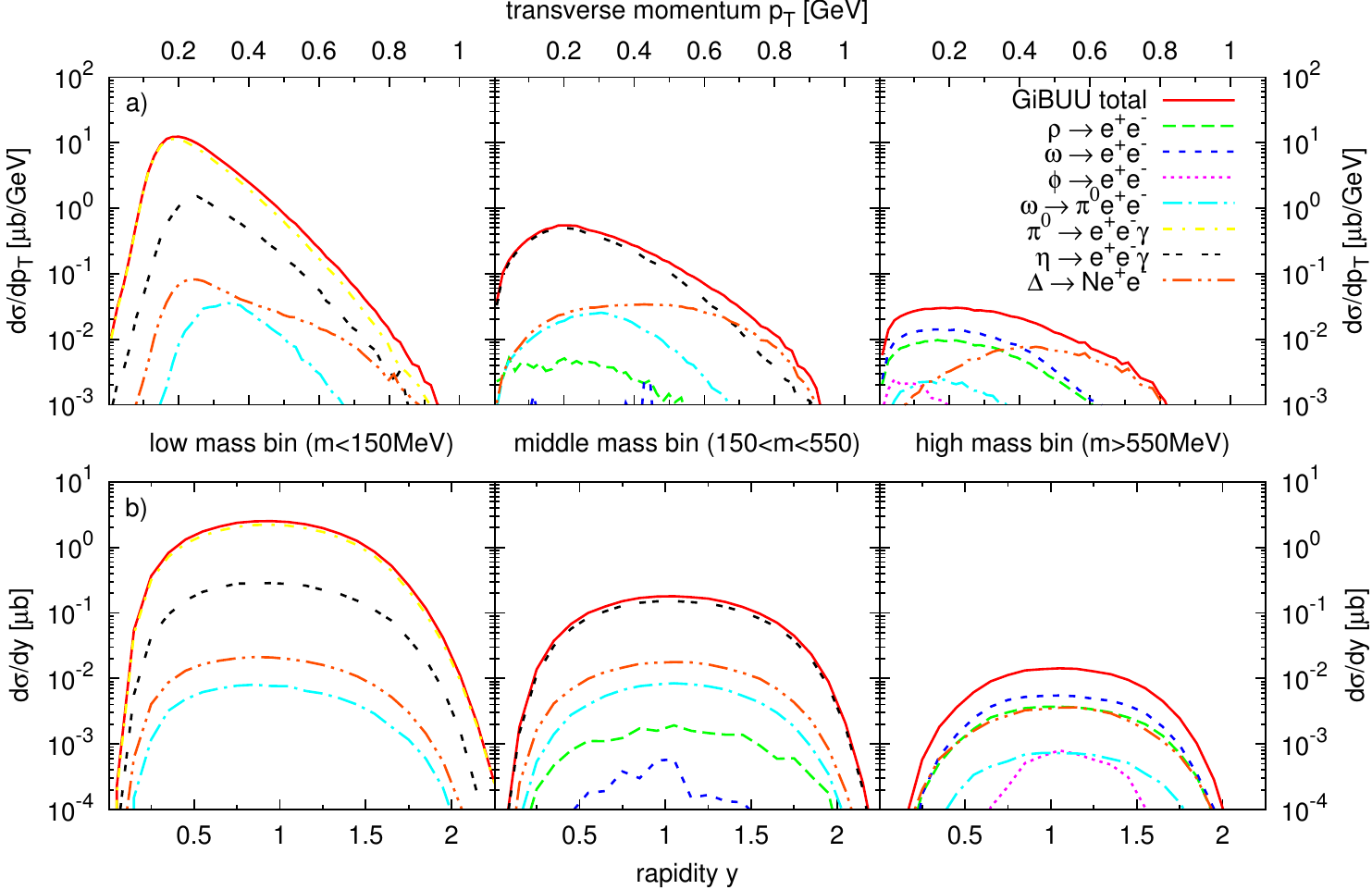}
  \end{center}
  \caption{(Color online) a) Transverse momentum and b) rapidity spectra of
    dilepton pairs for three mass bins (with $\Delta$ form factor).}
  \label{fig:pp_pt_y}
\end{figure}

In order to understand the underlying processes, it is not sufficient
to consider only the mass spectrum. Other observables can give further
insight into the reaction dynamics and can serve as a cross check for the
validation of theoretical models. We choose to examine the transverse momentum
and rapidity distribution in three different mass bins
(see fig.~\ref{fig:pp_pt_y}):

\begin{enumerate}
\item The low mass bin ($m < 150\MeV$) is clearly dominated by the
  $\pi^0$ Dalitz channel, with some admixture of the $\eta$ Dalitz.
\item The intermediate mass range of $150\MeV<m<550\MeV$ mostly
  contains contributions of the $\eta$ and $\Delta$ Dalitz decays.
\item The high mass bin ($550\MeV<m$) is where the vector mesons are
  situated.
\end{enumerate}

Distinguishing several mass bins is useful in order to separate the
contributions of different channels.
In all the three mass bins, we achieve a rather good agreement with
preliminary data (not shown) \cite{Rustamov_priv}, in transverse momentum as
well as rapidity. With \Pythia{}'s default parameters, the $p_T$ spectra show a
slight deviation from the data. The rapidity spectra turned out to be rather
insensitive to the details of the reaction dynamics and to most of the
parameters we modified. Their shape is mostly governed by the geometrical
acceptance of the detector, which we apply to our simulations via the HADES
acceptance filter (HAFT, version 2.0) \cite{hades,galatyuk_priv}.
Therefore, the agreement in the rapidity
spectra is mostly useful as a consistency check for the filtering
process. The overall agreement proves to be quite good, although there
are minor deviations at forward rapidity.

The asymmetry of the deviation indicates that the deviation is
indeed caused by acceptance filtering problems, since the unfiltered
physical distributions are symmetric around mid-rapidity, and any
asymmetries can only be introduced by the acceptance filter.

Within this level of agreement in the elementary p+p collisions, we
have a good baseline for studying in-medium effects in p+Nb, although
the issue of the $\Delta$ Dalitz form factor is not completely settled.

%%%%%%%%%%%%%%%%%%%%%%%%%%%%%%%%%%%%%%%%%%%%%%%%%%%%%%%%%%%%%%%%%%%%%%%%%%%%%%%

\section{Dilepton Spectra from p+Nb Collisions}

In p+Nb reactions there will be couple of additional effects,
compared to the elementary p+p reactions. First of all, the primary p+N
collisions will be nearly identical, apart from binding effects and
some Fermi smearing, but besides p+p also p+n collisions will play a
role.  Furthermore, the produced particles will undergo final state
interaction within the Nb nucleus, and processes like meson absorption
and regeneration may become important. The secondary collisions will
at average have lower energies than the primary N+N collisions,
therefore also the low-energy resonance part of the collision term
will be involved. Finally, also the vector meson spectral
functions may be modified in the nuclear medium.

\begin{figure}[t]
  \includegraphics[width=0.95\textwidth]{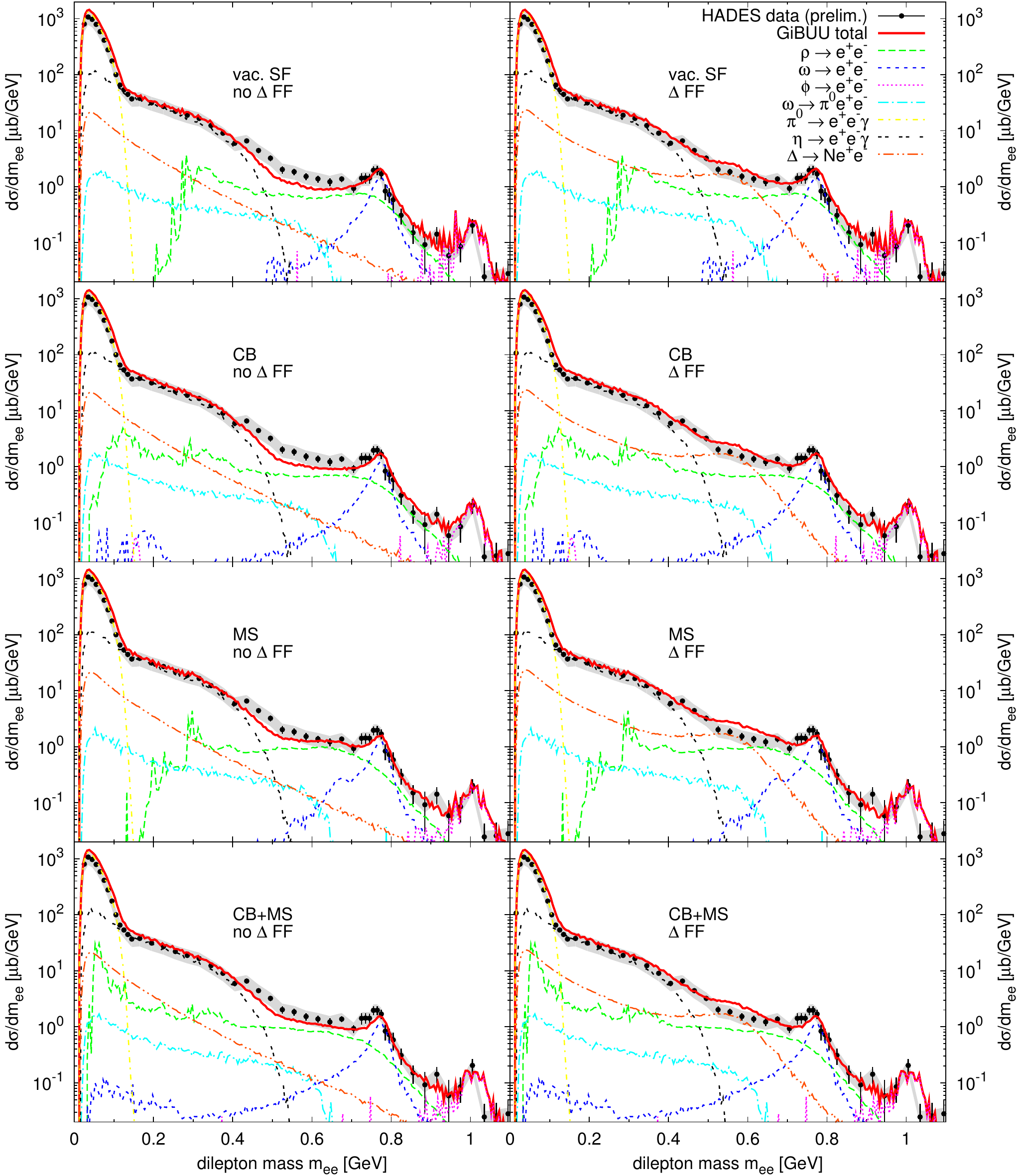}
  \caption{(Color online) Dilepton mass spectra for p+Nb@3.5\GeV. Left column:
    without $\Delta$ form factor. Right column: with $\Delta$ form factor.
    From top to bottom: vacuum spectral functions for the vector mesons,
    collisional broadening, $16\%$ mass shift, collisional broadening plus
    mass shift.}
  \label{fig:pNb_mass}
\end{figure}

Fig.~\ref{fig:pNb_mass} shows simulated dilepton spectra for p+Nb collisions at
3.5 \GeV in various scenarios, compared to the data from \cite{Weber:2011zz}.
The overall agreement is not quite as good as in the p+p case. Already in the
pion channel we slightly overestimate the data. This might have
various reasons: too little absorption or too much secondary pion
production in GiBUU, or even a normalization problem in the data. 

According to \cite{weber_phd}, the data have been normalized by
comparing charged pion spectra measured by HADES in p+Nb to those
measured by the HARP collaboration. However, the cross sections
obtained by HARP had to be extrapolated to the slightly different beam
energy and nuclear target of HADES. This procedure is responsible for
most of the systematic error of the data (roughly 28\%), which
is shown as a gray band in the figures.

Another striking feature of the p+Nb system is that the simulation
gets close to the data in the intermediate mass range, even without
any $\Delta$ form factor. Including the form factor will slightly
overshoot the data. It seems that most of the intermediate-mass gap
observed in p+p is filled up by low-mass $\rho$ mesons,
which presumably are produced in secondary collisions. Even in p+p
collisions, one might already get a similar effect by describing
$\rho$ meson production via resonance excitation, which could give
stronger contributions in the low-mass part of the $\rho$ spectral
function than \Pythia{}'s string fragmentation model.

\begin{figure}[t]
  \begin{center}
   \includegraphics[width=0.9\textwidth]{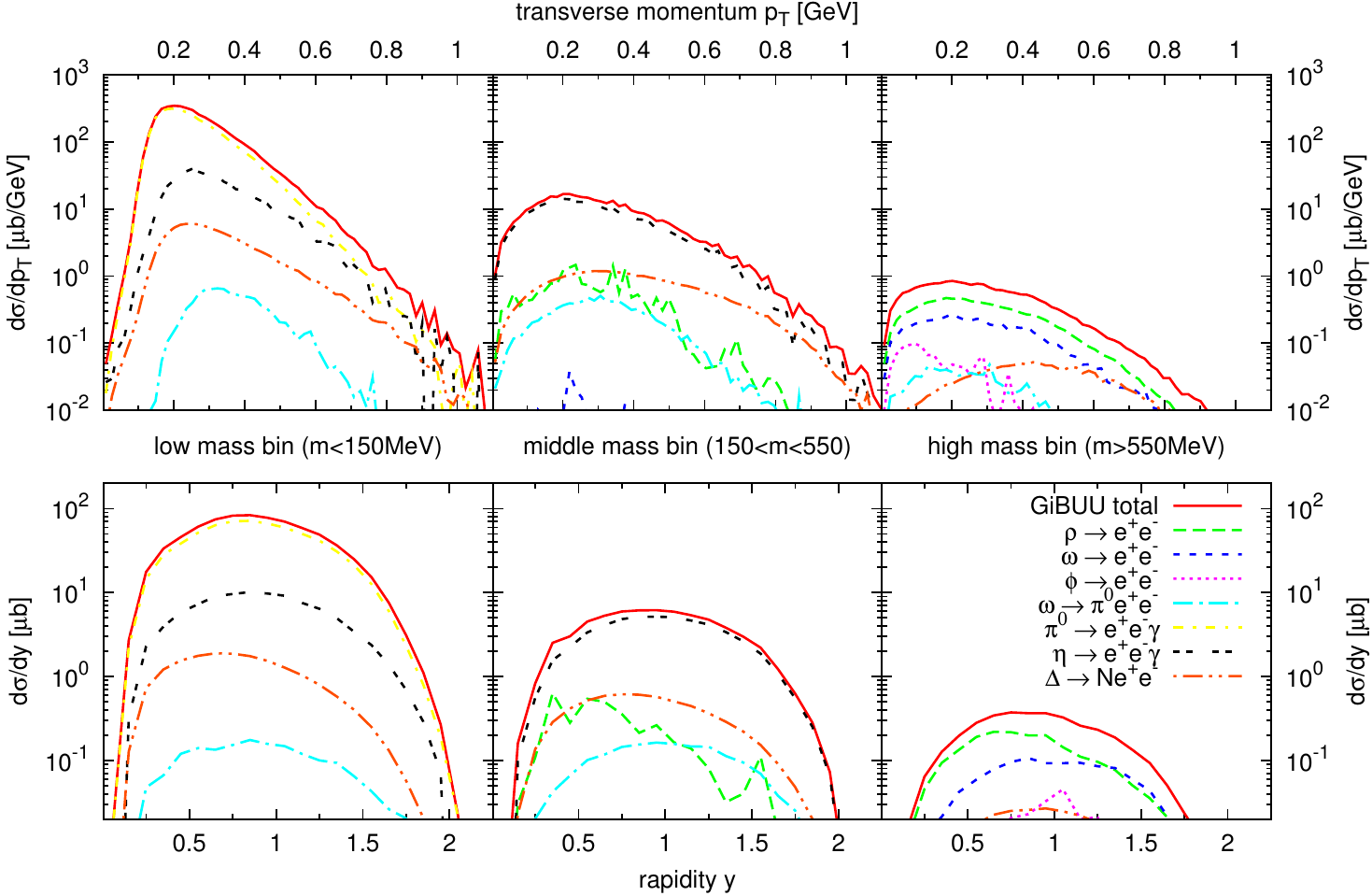}
  \end{center}
  \caption{(Color online) $p_T$ and rapidity spectra for three mass bins:
           $m<150\MeV$, $150\MeV<m<550\MeV$ and $550\MeV<m$.}
  \label{fig:pNb_pt_y}
\end{figure}

The mass spectrum above 500 \MeV can receive further modifications from
the inclusion of in-medium effects in the vector meson spectral
functions. Fig.~\ref{fig:pNb_mass} shows a few typical in-medium scenarios:
The first one includes a collisionally broadened in-medium width as described
in chapter \ref{sec:gibuu}, while the second one assumes a pole mass shift
according to
\begin{equation}
 m^*(\rho) = m_0\left(1-\alpha\frac{\rho}{\rho_0}\right) \; ,
\end{equation}
with a scaling parameter $\alpha = 16 \%$.
The third scenario combines both of these effects. The modifications
introduced by these scenarios are roughly on the same order of magnitude
as the systematic errors of the data. This fact, together with the
discrepancy in the $\pi^0$ channel and the uncertainty of the $\Delta$
form factor, presently disables us to draw any hard conclusion on vector meson
properties in cold nuclear matter from the HADES data.

The shape of the $p_T$ and rapidity spectra, depicted in
fig.~\ref{fig:pNb_pt_y}, show a similarly good agreement to preliminary data
(not shown) \cite{weber_priv} as in the p+p case. The shown $p_T$ and rapidity
spectra do neither include a $\Delta$ form factor nor any in-medium effects for
the vector mesons and are not significantly sensitive to such modifications.

%%%%%%%%%%%%%%%%%%%%%%%%%%%%%%%%%%%%%%%%%%%%%%%%%%%%%%%%%%%%%%%%%%%%%%%%%%%%%%%

\section {Conclusions}

We have shown that the HADES data from elementary p+p collisions at
3.5 \GeV kinetic beam energy can be described very well by the
\Pythia{} event generator with a few adjusted parameters, as
employed by the GiBUU model. It was found that the intermediate
mass region can be explained by including a VMD-like transition form
factor for the $\Delta$ Dalitz decay. Also the p+Nb reaction at the same beam
energy is reasonably well described by the GiBUU transport
model. However, a few discrepancies are left, which prevent us from
drawing final conclusions on in-medium effects at this point. We infer that the
HADES data at 3.5 \GeV can not be understood without fixing the open issue of
the $\Delta$ Dalitz form factor.

%%%%%%%%%%%%%%%%%%%%%%%%%%%%%%%%%%%%%%%%%%%%%%%%%%%%%%%%%%%%%%%%%%%%%%%%%%%%%%%

\appendix

\section{\Pythia{} Parameters}

GiBUU relies on the \Pythia{} event generator for elementary
collisions at c.m.~energies above 2.2 (2.6) \GeV for
meson-baryon (baryon-baryon) systems. \Pythia{}'s default parameters
are tuned to rather high energies, just like most of the available
parameter sets (called 'tunes' in \Pythia{}). To apply \Pythia{}
successfully to collisions at HADES energies, we adjusted a
couple of parameters. The following parameter set represents an
attempt to create a 'HADES tune' for \Pythia{}. A thorough
description of all the parameters can be found in the \Pythia{} manual
\cite{Sjostrand:2006za}.

\begin{table}[h]
  \begin{center}
  \begin{tabular}{|c|c|c|c|}
    \hline
    Parameter & Default & Professor                  & Our \\
	      & value   & tune \cite{Buckley:2009bj} & value \\
    \hline
    PARJ(11) & 0.5  & 0.31  & 0.15 \\
    PARJ(12) & 0.6  & 0.4   & 0.2 \\
    PARJ(21) & 0.36 & 0.313 & 0.25 \\
    PARJ(25) & 1.0  & 0.63  & 0.63 \\
    PARJ(26) & 0.4  & 0.12  & 0.12 \\
    PARP(91) & 2.0  &       & 0.25 \\
    \hline
  \end{tabular}
  \end{center}
  \caption{\Pythia{} parameters tuned to HADES energies.}
  \label{tab:Py_par}
\end{table}

The most prominent modification is the suppression of the vector meson
contributions by tuning down PARJ(11) and PARJ(12) by a
significant amount. PARJ(25) and PARJ(26) are the parameters for
suppressing the $\eta$ and $\eta'$ contribution, which we adopt from the
'Professor' tune \cite{Buckley:2009bj}. Moreover, the amount of
transverse momentum has to be adjusted via PARP(91) and PARJ(21),
which control the primordial $k_T$ and the $p_T$ from fragmentation,
respectively.

%%%%%%%%%%%%%%%%%%%%%%%%%%%%%%%%%%%%%%%%%%%%%%%%%%%%%%%%%%%%%%%%%%%%%%%%%%%%%%%

\acknowledgments

We thank the HADES collaboration for providing us with the data and the HADES
acceptance filter and for many fruitful discussions. Special thanks go to
Tetyana Galatyuk, Anar Rustamov and Michael Weber.
This work was supported by HGS-HIRe.

%%%%%%%%%%%%%%%%%%%%%%%%%%%%%%%%%%%%%%%%%%%%%%%%%%%%%%%%%%%%%%%%%%%%%%%%%%%%%%%

\bibliographystyle{JHEP}
\bibliography{references}

\providecommand{\href}[2]{#2}\begingroup\raggedright\begin{thebibliography}{10}

\bibitem{Leupold:2009kz}
S.~Leupold, V.~Metag, and U.~Mosel, {\it {Hadrons in strongly interacting
  matter}},  {\em Int.J.Mod.Phys.} {\bf E19} (2010) 147--224,
  [\href{http://xxx.lanl.gov/abs/0907.2388}{{\tt arXiv:0907.2388}}].

\bibitem{Wood:2008ee}
{\bf CLAS} Collaboration, M.~Wood {\em et.~al.}, {\it {Light Vector Mesons in
  the Nuclear Medium}},  {\em Phys.Rev.} {\bf C78} (2008) 015201,
  [\href{http://xxx.lanl.gov/abs/0803.0492}{{\tt arXiv:0803.0492}}].

\bibitem{Naruki:2005kd}
M.~Naruki, H.~Funahashi, Y.~Fukao, M.~Kitaguchi, M.~Ishino, {\em et.~al.}, {\it
  {Experimental signature of the medium modification for rho and omega mesons
  in 12-GeV p + A reactions}},  {\em Phys.Rev.Lett.} {\bf 96} (2006) 092301,
  [\href{http://xxx.lanl.gov/abs/nucl-ex/0504016}{{\tt nucl-ex/0504016}}].

\bibitem{:2008yh}
{\bf HADES} Collaboration, I.~Frohlich {\em et.~al.}, {\it {Meson and
  di-electron production with HADES}},  {\em Int.J.Mod.Phys.} {\bf A24} (2009)
  317--326, [\href{http://xxx.lanl.gov/abs/0809.2764}{{\tt arXiv:0809.2764}}].

\bibitem{Nanova:2010tq}
{\bf CBELSA/TAPS} Collaboration, M.~Nanova {\em et.~al.}, {\it {Photoproduction
  of $\omega$ mesons on nuclei near the production threshold}},  {\em
  Eur.Phys.J.} {\bf A47} (2011) 16,
  [\href{http://xxx.lanl.gov/abs/1008.4520}{{\tt arXiv:1008.4520}}].

\bibitem{gibuu}
``{GiBUU website}.''
\newblock \href{http://gibuu.physik.uni-giessen.de}{\tt
  http://gibuu.physik.uni-giessen.de}.

\bibitem{Cassing:1999wx}
W.~Cassing and S.~Juchem, {\it Semiclassical transport of particles with
  dynamical spectral functions},  {\em Nucl. Phys. A} {\bf 665} (2000)
  377--400, [\href{http://xxx.lanl.gov/abs/nucl-th/9903070}{{\tt
  nucl-th/9903070}}].

\bibitem{Teis:1996kx}
S.~Teis, W.~Cassing, M.~Effenberger, A.~Hombach, U.~Mosel, {\em et.~al.}, {\it
  {Pion production in heavy ion collisions at SIS energies}},  {\em Z.Phys.}
  {\bf A356} (1997) 421--435,
  [\href{http://xxx.lanl.gov/abs/nucl-th/9609009}{{\tt nucl-th/9609009}}].

\bibitem{pythia}
``{Pythia website}.''
\newblock \href{http://projects.hepforge.org/pythia6}{\tt
  http://projects.hepforge.org/pythia6}.

\bibitem{Sjostrand:2006za}
T.~Sjostrand, S.~Mrenna, and P.~Skands, {\it {PYTHIA 6.4 Physics and Manual}},
  {\em JHEP} {\bf 05} (2006) 026,
  [\href{http://xxx.lanl.gov/abs/hep-ph/0603175}{{\tt hep-ph/0603175}}].

\bibitem{Gallmeister:2009ht}
K.~Gallmeister and U.~Mosel, {\it {Production of charged pions off nuclei with
  3...30 GeV incident protons and pions}},  {\em Nucl. Phys.} {\bf A826} (2009)
  151--160, [\href{http://xxx.lanl.gov/abs/0901.1770}{{\tt arXiv:0901.1770}}].

\bibitem{effe_phd}
M.~Effenberger, {\em Eigenschaften von Hadronen in Kernmaterie in einem
  vereinheitlichten Transportmodell}.
\newblock PhD thesis, Justus-Liebig-Universit\"at Gie\ss{}en, 1999.
\newblock available online at \url{http://theorie.physik.uni-giessen.de/}.

\bibitem{Nakamura:2010zzi}
{\bf Particle Data Group} Collaboration, K.~Nakamura {\em et.~al.}, {\it
  {Review of particle physics}},  {\em J.Phys.G} {\bf G37} (2010) 075021.

\bibitem{Berlowski:2008zz}
M.~Berlowski, C.~Bargholtz, M.~Bashkanov, D.~Bogoslawsky, A.~Bondar, {\em
  et.~al.}, {\it {Measurement of eta meson decays into lepton-antilepton
  pairs}},  {\em Phys.Rev.} {\bf D77} (2008) 032004.

\bibitem{Browder:1997eu}
{\bf CLEO} Collaboration, T.~Browder {\em et.~al.}, {\it {A New upper limit on
  the decay eta --> e+ e-}},  {\em Phys.Rev.} {\bf D56} (1997) 5359--5365,
  [\href{http://xxx.lanl.gov/abs/hep-ex/9706005}{{\tt hep-ex/9706005}}].

\bibitem{Landsberg:1986fd}
L.~G. Landsberg, {\it {Electromagnetic Decays of Light Mesons}},  {\em Phys.
  Rept.} {\bf 128} (1985) 301--376.

\bibitem{spruck_phd}
B.~Spruck, {\em Optimierung des Pionenstrahls zum HADES Detektor und Bestimmung
  des Eta-Formfaktors in Proton-Proton Reaktionen bei 2.2 GeV}.
\newblock PhD thesis, Justus-Liebig-Universit\"at Gie\ss{}en, 2008.
\newblock http://geb.uni-giessen.de/geb/volltexte/2008/6667/.

\bibitem{Terschluesen:2010ik}
C.~Terschlusen and S.~Leupold, {\it {Electromagnetic transition form factors of
  light vector mesons}},  {\em Phys.Lett.} {\bf B691} (2010) 191--201,
  [\href{http://xxx.lanl.gov/abs/1003.1030}{{\tt arXiv:1003.1030}}].

\bibitem{Bratkovskaya:1996qe}
E.~L. Bratkovskaya and W.~Cassing, {\it {Dilepton production from AGS to SPS
  energies within a relativistic transport approach}},  {\em Nucl. Phys.} {\bf
  A619} (1997) 413--446, [\href{http://xxx.lanl.gov/abs/nucl-th/9611042}{{\tt
  nucl-th/9611042}}].

\bibitem{:2009wb}
{\bf NA60} Collaboration, R.~Arnaldi {\em et.~al.}, {\it {Study of the
  electromagnetic transition form-factors in eta --> mu+ mu- gamma and omega
  --> mu+ mu- pi0 decays with NA60}},  {\em Phys.Lett.} {\bf B677} (2009)
  260--266, [\href{http://xxx.lanl.gov/abs/0902.2547}{{\tt arXiv:0902.2547}}].

\bibitem{Krivoruchenko:2001hs}
M.~I. Krivoruchenko and A.~Faessler, {\it {Comment on Delta radiative and
  Dalitz decays}},  {\em Phys. Rev.} {\bf D65} (2002) 017502,
  [\href{http://xxx.lanl.gov/abs/nucl-th/0104045}{{\tt nucl-th/0104045}}].

\bibitem{Wan:2005ds}
Q.~Wan and F.~Iachello, {\it {A unified description of baryon electromagnetic
  form factors}},  {\em Int.J.Mod.Phys.} {\bf A20} (2005) 1846--1849.

\bibitem{Rustamov:2010zz}
{\bf HADES} Collaboration, A.~Rustamov, {\it {Inclusive meson production in
  3.5-GeV p p collisions studied with the HADES spectrometer}},  {\em AIP
  Conf.Proc.} {\bf 1257} (2010) 736--740.

\bibitem{Rustamov_priv}
A.~Rustamov, ``private communications.''

\bibitem{hades}
``{HADES website}.''
\newblock \href{http://www-hades.gsi.de}{\tt http://www-hades.gsi.de}.

\bibitem{galatyuk_priv}
T.~Galatyuk, ``private communications.''

\bibitem{Weber:2011zz}
{\bf HADES Collaboration} Collaboration, M.~Weber, {\it {Dielectron
  spectroscopy in cold nuclear matter}},  {\em Int.J.Mod.Phys.} {\bf A26}
  (2011) 737--740.

\bibitem{weber_phd}
M.~Weber, {\em Dielektronen-Spektroskopie in kalter Kernmaterie}.
\newblock PhD thesis, Technische Universit\"at M\"unchen, 2001.
\newblock available online at \url{http://mediatum.ub.tum.de/node?id=1007264}.

\bibitem{weber_priv}
M.~Weber, ``private communications.''

\bibitem{Buckley:2009bj}
A.~Buckley, H.~Hoeth, H.~Lacker, H.~Schulz, and J.~E. von Seggern, {\it
  {Systematic event generator tuning for the LHC}},  {\em Eur. Phys. J.} {\bf
  C65} (2010) 331--357, [\href{http://xxx.lanl.gov/abs/0907.2973}{{\tt
  arXiv:0907.2973}}].

\end{thebibliography}\endgroup

\end{document}